\documentclass{aa}
\usepackage{amssymb}
\usepackage{amsmath}
\usepackage{txfonts}
\usepackage{graphicx}
\usepackage{natbib}
\usepackage{rotating}

\begin{document}

\title{Long Term Variability of Cyg~X-1}
\subtitle{II. The rms-Flux Relation}

\author{T.~Gleissner\inst{1} \and J.~Wilms\inst{1} \and
   K.~Pottschmidt\inst{2,3} \and P.~Uttley\inst{4} \and 
   M.A.~Nowak\inst{5} \and R.~Staubert\inst{1}}
\institute{Institut f\"ur Astronomie und Astrophysik -- Abt.~Astronomie,
   Universit\"at T\"ubingen, Sand 1, 72076 T\"ubingen, Germany
\and Max-Planck-Institut f\"ur extraterrestrische
   Physik, Postfach 1312, 85748 Garching, Germany
\and INTEGRAL Science
   Data Centre, Chemin d'\'Ecogia 16, 1290 Versoix, Switzerland
\and School of Physics and Astronomy, University of Southampton,
   Southampton S017~1BJ, UK
\and MIT-CXC, NE80-6077, 77 Massachusetts Ave., Cambridge, MA 02139, USA}

\mail{T.~Gleissner (gleiss@astro.uni-tuebingen.de)}
\titlerunning{Long Term Variability of Cygnus~X-1: II.}
\authorrunning{T.~Gleissner et al.}
\date{Received $<$date$>$ / Accepted $<$date$>$ }

\abstract{We study the long term evolution of the relationship between the
  root mean square (rms) variability and flux (the ``rms-flux relation'')
  for the black hole Cygnus~X-1 as monitored from 1996 to 2003 with the
  Rossi X-ray Timing Explorer (\textsl{RXTE}). We confirm earlier results by
  \citet{uttley:01} of a linear relationship between rms and flux in the
  hard state on time scales $>5$\,s reflecting in its slope the
  fractional rms variability. We demonstrate the perpetuation of the linear
  rms-flux relation in the soft and the intermediate state. The existence
  of a non-zero intercept in the linear rms-flux relation argues for two
  lightcurve components, for example, one variable and one non-variable
  component, or a possible constant rms component. The relationship between
  these two hypothesized components can be described by a fundamental
  dependence of slope and intercept at time scales $\lesssim 10$\,ksec with
  \emph{long term} averages of the flux. 
   }
\maketitle

\section{Introduction}\label{sec:intro}

Galactic Black holes (BHs) are generally found in two major states which
are mainly characterized by their X-ray and radio spectral and timing
properties \citep[see, e.g.,][and references therein]{nowak:02a}. The
``hard state'', occuring at luminosities $L\lesssim 0.05\,L_{\text{Edd}}$,
where $L_{\text{Edd}}$ is the Eddington luminosity, is characterized by
$\sim$30\% root mean square (rms) variability, a hard X-ray spectrum that
can be roughly described by an exponentially cutoff power law, and radio
emission. The hard state energy spectrum can physically be modelled by
thermal Comptonization of cool seed photons in a hot electron plasma
\citep{thorne:75,shapiro:76,sunyaev:79a,dove:97c}. At higher luminosities,
BHs are seen in the ``soft state'' with a thermal X-ray spectrum with a
characteristic temperature of at most a few keV, X-ray variability of only
a few percent rms, and no observed radio emission \citep{fender:02a}. The
thermal X-ray spectrum in the soft state is generally associated with a
geometrically thin, optically thick accretion disk
\citep{novikov:73a,shakura:73,gierlinski:98a}. In addition to these two
canonical states, further states have been identified which are
characterized either by an even larger luminosity than in the soft state
(the ``very high state''), or by variability and X-ray spectral properties
which are mostly intermediate between the hard and the soft state
\citep[the ``intermediate state'';][]{belloni:96}.

The analysis of the short term X-ray variability of BHs is especially
well suited for gaining better insight into the physical processes at
work near the compact object. In order to analyze this behavior in a
systematic way, in 1997 we initiated a long term monitoring campaign
of the canonical hard state BH Cygnus~X-1. During the campaign, which
is still ongoing, Cyg~X-1 is observed at $\sim$7\,d to $\sim$14\,d
intervals simultaneously with the Rossi X-ray Timing
Explorer (\textsl{RXTE}), the Ryle radio telescope, and several optical
instruments.  These observations enable us to study the evolution of
the observational properties of Cyg~X-1 (e.g., X-ray spectrum and
timing parameters, radio flux) and thus the source history on
time scales longer than the orbital time scale of $\sim 5.6$\,days.

Earlier results from our campaign have been reported by
\citet{pottschmidt:00a} and by \citet[][hereafter paper~I]{pottschmidt:03},
where we studied the evolution of the Fourier-frequency dependent X-ray
time lags and of the power spectral density (PSD), respectively. Following
\citet{nowak:00a}, in paper~I we were able to show that for a very wide
range of source parameters, the PSD of Cyg~X-1 is well described by the
superposition of broad Lorentzian functions. Such a description is an
alternative to the more common description of the PSD in terms of broken
power laws \citep{klis:95} or shot-noise profiles \citep[][and references
therein]{lochner:91}.  During changes of the source into the ``intermediate
state'' or the ``soft state'', the characteristic frequency of these
components shifts towards higher frequencies in a systematic manner, some
of the Lorentzians change their strength or vanish, and the X-ray time lag
increases in the frequency band dominated by those Lorentzians which become
stronger \citep[][paper~I]{pottschmidt:00a}. Once the ``soft state'' is
reached, the X-ray time lags are again similar to those seen during the
hard state.

In paper~I we interpreted these results in terms of models invoking
damped oscillations in the accretion disk and/or the Comptonizing
plasma \citep{dimatteo:99a,psaltis:00a,churazov:01a,nowak:02b}.
Because of the complicated physics of the system ``accretion disk --
accretion disk corona -- radio outflow'', none of these models are
fully self-consistent and thus none of the models describes all
observed properties of Cyg~X-1.  It is our hope that additional
empirical data for the hard state behavior might help to further
constrain these different emission models of BHs.

Recently, \citet{uttley:01} found a remarkable linear relationship between
the rms amplitude of broadband noise variability and flux in the X-ray
lightcurves of neutron star and BH X-ray binaries and active galactic
nuclei (AGN). This linear relation between flux and rms variability has an
offset on the flux axis, leading to the suggestion that the lightcurves
of X-ray binaries are made from at least two components: one component with
a linear dependence of rms variability on flux, and one component which
contributes a constant rms to the lightcurve or does not vary at all. The
fact that the same rms-flux relation was found in the lightcurves of active
galactic nuclei suggests that this behavior is intrinsic to compact
accreting systems.

In the case of X-ray binaries, the rms-flux relation has
until now only been investigated for single observations of
SAX~J1808.4$-$3658 and for a few hard state observations of Cyg~X-1.  It is
important to discover whether this relation applies generally for many
observations of a given object and across a range of states.  Therefore in
this paper we investigate the evolution and behavior of the rms-flux
relation in Cyg X-1 using the \textsl{RXTE} monitoring data. Preliminary
results have already been presented elsewhere \citep{gleissner:02a}.

The remainder of this paper is structured as follows. In
Sect.~\ref{sec:data} we describe the data and the computation of the rms
variability. We then present our results on the rms-flux relation
(Sect.~\ref{sec:results}). We show that the rms-flux relation is valid
throughout all spectral states, for all energies and on all time scales. It
is demonstrated how mean flux, $\langle F\rangle$, and the relation fit
parameters slope, $k$, and intercept on the flux axis, $C$, map out a
fundamental plane in the hard state. After analyzing spectral dependences,
we examine the rms-flux relation on short time scales by calculating the
variability of the rms variability. We end the paper with a discussion of
the physical interpretations, and explore the meaning of the long term
rms-flux relation as well as of the hard state $k$-$C$-$\langle F\rangle$
fundamental plane (Sect.~\ref{sec:disc}).

\section{Observations and data analysis}\label{sec:data}

\subsection{Data extraction}
\begin{table}
\caption{Definition of the energy
  bands and data modes.}\label{tab:energybands} 
\noindent
\begin{tabular}{l@{\hspace*{2mm}}l@{\hspace*{2mm}}l@{\hspace*{2mm}}l@{\hspace*{2mm}}l@{\hspace*{2mm}}l}
        & band 1 & band 2 & band 3 & band 4 & band 5\\
\hline\hline
\multicolumn{6}{l}{\textsl{PCA} Epoch~3: data taken until 1999 March 22}\\
\hline
\multicolumn{6}{l}{\rule{0mm}{4mm}P10236$^a$} \\
channels    & 0--9  & 10--15 & 16--21 & 22--35 & 36--189 \\
E [keV]& $\sim$2--3.7  & 3.7--5.8 & 5.8--8.0 & 8.0--13.0 & 13.0--73.5 \\
\hline
\multicolumn{6}{l}{\rule{0mm}{4mm}P10241$^b$} \\
channels    & 0--10  & 11--16 & 17--22 & 23--35 & 36--189 \\
E [keV]& $\sim$2--4.1  & 4.1--6.2 & 6.2--8.3 & 8.3--13.0 & 13.0--73.5 \\
\hline
\multicolumn{6}{l}{\rule{0mm}{4mm}P10257$^c$} \\
channels    & 0--13  & 14--23 & 24--35 & 36--249 &  \\
E [keV]& $\sim$2--5.1  & 5.1--8.7 & 8.7--13.0 & 13.0--100.4 &  \\
\hline
\multicolumn{6}{l}{\rule{0mm}{4mm}P10412$^d$} \\
channels    & 0--9  & 10--15 & 16--21 & 22--35 & 36--174 \\
E [keV]& $\sim$2--3.7  & 3.7--5.8 & 5.8--8.0 & 8.0--13.0 & 13.0--67.1 \\
\hline
\multicolumn{6}{l}{\rule{0mm}{4mm}P10512$^e$} \\
channels    & 0--9  & 10--15 & 16--21 & 22--35 & 36--174 \\
E [keV]& $\sim$2--3.7  & 3.7--5.8 & 5.8--8.0 & 8.0--13.0 & 13.0--67.1 \\
\hline
\multicolumn{6}{l}{\rule{0mm}{4mm}P30157$^f$} \\
channels    & 0--10  & 11--16 & 17--22 & 23--35 & 36--49 \\
E [keV]& $\sim$2--4.1  & 4.1--6.2 & 6.2--8.3 & 8.3--13.0 & 13.0--18.1 \\
\hline
\multicolumn{6}{l}{\rule{0mm}{4mm}P40099-01 to P40099-05$^g$} \\
channels    & 0--10  & 11--16 & 17--22 & 23--35 & 36--189 \\
E [keV]& $\sim$2--4.1  & 4.1--6.2 & 6.2--8.3 & 8.3--13.0 & 13.0--73.5 \\
\hline\hline
\multicolumn{6}{l}{\textsl{PCA} Epoch~4: data taken after 1999 March 22}\\
\hline
\multicolumn{6}{l}{\rule{0mm}{4mm}P40099-06 to P40099-28, P50110, P60090, P70414$^h$} \\
channels    & 0--10 & 11--13 & 14--19 & 20--30 & 36--159 \\
E [keV]& $\sim$2--4.6  & 4.6--5.9 & 5.9--8.4 & 8.4--13.1 & 15.2--71.8\\
\hline
\end{tabular}
\vspace{2mm}

\footnotesize

${}^{a}$ data modes B\_4ms\_8A\_0\_35\_H and E\_62us\_32M\_36\_1s\\  
${}^{b}$ data modes B\_2ms\_8B\_0\_35\_Q and E\_125us\_64M\_36\_1s\\
${}^{c}$ data mode E\_4us\_4B\_0\_1s\\
${}^{d}$ data modes B\_4ms\_8A\_0\_35\_H and E\_62us\_32M\_36\_1s\\
${}^{e}$ data modes B\_4ms\_8A\_0\_35\_H and E\_16us\_16B\_36\_1s\\
${}^{f}$ data mode B\_16ms\_46M\_0\_49\_H\\
${}^{g}$ data modes B\_2ms\_8B\_0\_35\_Q and E\_125us\_64M\_36\_1s\\
${}^{h}$ data modes B\_2ms\_8B\_0\_35\_Q and E\_125us\_64M\_36\_1s\\

\end{table}

Most of the data analyzed in this paper are the lightcurves that
we already used in paper~I of this series. We therefore only give a
brief summary of the \textsl{RXTE} campaign and the data extraction issues and
refer to paper~I for the details.

The data analyzed here spans the time from 1996 until early 2003, covering
our \textsl{RXTE} monitoring programs since \textsl{RXTE} Announcement of
Opportunity 3 (AO3, 1998) through \textsl{RXTE}'s AO6 (2002/2003).
Additionally, other public \textsl{RXTE} data from \textsl{RXTE} AO1 and
AO7 were used.  Typical exposure times were 3\,ksec in 1998, and 10\,ksec
since then.  After screening the data for episodes of increased background
(see paper~I), we extracted lightcurves with a resolution of $2^{-6}$\,s
($\sim$16\,ms; AO3) or $2^{-8}$\,s ($\sim$4\,ms; other data) using data
from the \textsl{RXTE} Proportional Counter Array \citep{jahoda:96} and the
standard \textsl{RXTE} data analysis software, \texttt{HEASOFT},
Version~5.0. We extracted lightcurves in 5 different energy bands
(Table~\ref{tab:energybands}). Due to integer overflows in the binned data
modes, high soft X-ray count rates can lead to a distortion of the rms-flux
relation for the lowest energy band (see Appendix~A for further
discussion). We therefore use only data from energy bands~2--5 which are
not affected. As the total number of operational proportional counter units
(\textsl{PCU}s) was often not constant during an observation, separate
lightcurves were generated for each of the different \textsl{PCU} combinations.
These lightcurves were analyzed separately. To facilitate comparisons,
unless noted otherwise we normalize all data to one \textsl{PCU}.

Table~\ref{tab:energybands} shows that the energy limits of the
high-resolution lightcurves, which were used for the computation of the
rms-flux relation, in parts differ significantly. In order to enable
comparisons, we normalize all observations to the energy limits of epoch 4
(P40099-06/28, P50110, P60090, P70414). We calculate a flux correction factor to
the epoch~3 observations by integrating the count rates of the
corresponding spectrum channels of each observation. This factor has to be
applied also to all values that are dependent on flux, e.g. $C$, $\sigma$, etc.

\subsection{Computation of the rms variability}\label{sec:calc}
To determine the rms variability as a function of source flux, we first cut
the observed lightcurve into segments of 1\,s duration for which the source
count rate is determined.  These segments are then assigned to their
respective flux bins. In general, 40 linearly spaced flux bins were chosen.
In the final computation, only flux bins containing at least 20 segments
were retained.  We note that short time segments of 1\,s duration are
necessary to study the characteristic variability of the source which is
dominated by frequencies above $10^{-1}$\,Hz.  Thus using a 1\,s segment
size allows us to measure rms-flux relations which sample a broad range of
fluxes, while also measuring rms with high signal-to-noise, since the
minimum frequency of variability sampled within each segment is well below
the frequency where photon counting noise dominates.

\begin{figure}
\resizebox{\hsize}{!}{\includegraphics{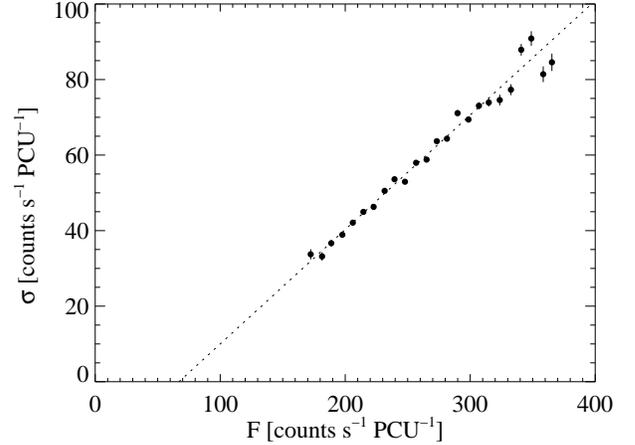}}
\caption{Relation between the \textsl{PCA} count rate, $F$, and the rms
  variability, $\sigma$, for the observation of 1998 July~02, showing the
  typical hard state characteristics. The data are extracted from energy
  band 2 (4.1--6.2\,keV). The dotted line is the best-fitting linear model
  to the data (see text), the error bars are at the 1$\sigma$ level.}
\label{fig:h4744f1}
\end{figure}

Next for each of the lightcurve segments in each of the
flux bins, the rms variability is determined for the Fourier frequency
interval of interest.  We first compute the PSDs of the individual
lightcurve segments \citep{nowak:98b,nowak:00a}, using the PSD
normalization where the integrated PSD equals the squared fractional
rms variability (i.e., the fractional variance) of the lightcurve
\citep{belloni:90b,miyamoto:92}.  We then average all the PSDs
obtained for segments in the same flux bin, and bin the averaged PSD
over the frequency range of interest (for this paper we use the range
from 1\,Hz to 32\,Hz) to yield the average power density in that
frequency range, $\langle P \rangle$.  Then, the absolute rms
variability of the source in the frequency band of interest, $\sigma$,
is obtained according to
\begin{equation}
\sigma = \left[\left(\langle P \rangle-c_{\text{Poisson}}\right) \cdot
  \Delta f  \right]^{1/2} F
\end{equation}
where $c_{\text{Poisson}}$ is the Poisson noise level due to photon
counting statistics, $\Delta f$ is the width of the frequency band of
interest, and where $F$ is the source count rate of the flux bin
(normalized to one \textsl{PCU}). Note that in the comparatively low frequency
range considered here, the detector deadtime does not need to be taken
into account when determining the Poisson noise level.  The
statistical uncertainty of the average PSD value, $\Delta \langle P
\rangle$, is determined from the periodogram statistics
\citep{klis:89a}
\begin{equation}
\label{eq:rmsuncert}
\Delta \langle P \rangle = \frac{\langle P \rangle}{\sqrt{MW}}
\end{equation}
where $M$ is the number of segments used in determining the average
rms-value and $W$ is the number of Fourier frequencies in the frequency
range of interest. The uncertainty of $\sigma$, $\Delta(\sigma)$, is then
computed using the standard error propagation formulae.  Note that, since
we measure $\sigma$ from the average PSD in each flux bin, our method
differs from that of \citet{uttley:01}, who first calculated a Poisson
noise-subtracted rms for each individual segment before binning up to
obtain $\sigma$ in each flux bin. \citet{uttley:01} computed the
uncertainty of $\sigma$ by calculating standard uncertainties from the
scatter of the rms values in each flux bin. This method has the
disadvantage that, when the variance due to the source is small compared to
the variance due to Poisson noise, subtracting the expected Poisson noise
level can sometimes lead to negative variances due to the intrinsic
variations in the realized noise level.  Our revised method avoids this
problem by averaging the segment PSDs before noise is subtracted, thus
reducing the intrinsic variations in the noise level.

\subsection{rms versus flux}\label{subsec:rmsvsflux}
Fig.~\ref{fig:h4744f1} shows the rms versus flux relationship for a
typical hard state observation, taken on 1998 July~02, in energy band 2
(4.1--6.2\,keV). To describe this linear relationship, we model it with a
function of the form \citep{uttley:01}
\begin{equation}
\label{eq:fitc}
\sigma = k \left(F-C\right)
\end{equation}
where the slope, $k$, and intercept, $C$, are the fit parameters.  This
linear model fits the general shape of the rms-flux relation of the example
reasonably well, although the large $\chi^2$ value of 102.1 for 22 degrees
of freedom implies that there is some weak intrinsic scatter. The
best-fitting model parameters for the example of
Fig.~\ref{fig:h4744f1} are $k=0.30\pm0.01$ and $C=67\pm
5\,\text{counts}\,\text{s}^{-1}\,\text{PCU}^{-1}$ (unless otherwise noted,
uncertainties are at the 90\% confidence level for two interesting
parameters).

In order to interpret the parameters of Eq.~\eqref{eq:fitc} it is useful to
use a toy model in which the observed source flux, $F_\text{obs}(t)$, is
written as the sum of two components, 
\begin{equation}\label{eq:sum}
F_\text{obs}(t) = F_\text{const} + F_\text{var}(t)
\end{equation}
where $F_\text{const}$ is a component of the lightcurve which is assumed to
be non-varying on the time scales considered here, and where
$F_\text{var}(t)$ is a component which is variable and thus responsible for
the observed rms variability. In the toy model of Eq.~\eqref{eq:sum} the
interpretation of the slope $k$ of the rms-flux relationship is
straightforward: it is the fractional rms of $F_\text{var}(t)$.
Furthermore, the intercept of the rms-flux relation on the flux axis, $C$,
is the count rate of $F_\text{const}$. It is obtained by extrapolating
$F_\text{var}\rightarrow 0$ in Eq.~\eqref{eq:sum}.

While the decomposition of Eq.~\eqref{eq:sum} is the most straightforward,
it is not fully unique. As we will show later, for the majority of our
observations $C>0$, however, there are several observations in which $C<0$.
In these cases the above interpretation clearly breaks down (there are no
negative fluxes) and it makes more sense to write Eq.~\eqref{eq:fitc} as
\begin{equation}\label{eq:sigma0}
\sigma = k F + \sigma_0
\end{equation}
In this formulation of the rms-flux relationship we have introduced
$\sigma_0$, an excess rms present in the lightcurve, which is positive
whenever $C<0$. Note that for $\sigma_0>0$ Eq.~\eqref{eq:sigma0} predicts
$\sigma \rightarrow \sigma_0$ for $F\rightarrow 0$, i.e., the extrapolation
of the rms-flux relationship to flux zero predicts that there is still
variability present, which is clearly unphysical. We do not consider this
behavior a problem, however, as the rms-flux relationship is measured only
for $F\gg 0$. We deem it reasonable to assume that for $F\simeq 0$ other
variability processes which do not obey the linear rms-flux relationship
will become important and force $\sigma\rightarrow 0$ for $F\rightarrow 0$.
Finally, we note that for $\sigma_0<0$ the observed rms is reduced with
respect to a linear rms-flux relationship with $\sigma_0=0$. This is
possible, e.g., if a constant flux component, $F_\text{const}$, is present
in the lightcurve (the majority of our fits imply reductions of about 25\%
in rms with respect to $\sigma_0=0$).

Observationally, degeneracies in the fit prevent us from distinguishing
between the two interpretations of the rms-flux relationship. Since most of
the observations of Cyg~X-1 presented in the following have $C>0$, and to
enable the comparison with other analyses, we have chosen to use
Eq.~\eqref{eq:fitc} as our fit function. We ask the reader, however, to
keep the second interpretation of the rms-flux relationship in mind.

The best test of linearity is a $\chi^2$ test of whether a straight
line fits the data. On the other hand, it is usually very difficult to
fit a simple model to variability properties because of the scatter in
the data points. It may be misleading to discard an observation as
being `non-linear' just because a straight line does not supply a good
$\chi^2$ fit. We found that Kendall's rank correlation coefficient,
which tests whether a correlation is monotonic, serves as a good
indicator of linearity in the sense that the bulk of points follows a
straight trend, while allowing for some points, predominantly at high fluxes
(cf. Fig.~\ref{fig:h4744f1}), to diverge.

Therefore we choose to characterize the goodness of the linear correlation
between $N$ data points of $\sigma$ and $F$ using Kendall's rank
correlation coefficient, $\tau$, defined by
\begin{equation}
\label{eq:tau}
\tau = \frac{S}{N (N-1)/2}
\end{equation}
Here $S$ is the sum of scores which are determined if we assign to
each of the $N \left(N-1\right)/2$ possible pairs
$[(\sigma_{\text{i}}/F_{\text{i}}),(\sigma_{\text{j}}/F_{\text{j}})]$
of the sample
$(\sigma_{1}/F_{1}),\ldots,(\sigma_{\text{N}}/F_{\text{N}})$ a score
of $+1$ or $-1$ depending on whether their ranks are in the same order
or in the opposite order on the $\sigma$ and $F$ axes, respectively
\citep{keeping:62}.

The correlation in Fig.~\ref{fig:h4744f1}, e.g., has $\tau = 0.95$
(the range of the correlation coefficient is $-1 \le \tau \le 1$, with a
perfect correlation being indicated by $\tau=1$). For the remainder of this
paper we set a minimum threshold of $\tau_{\text{min}}=0.9$ below which we
do not accept the hypothesis of a linear rms-flux relationship. Given this
criterion, the fit with this simple linear function works for the majority
of all observations (201 out of 220).  With a few exceptions, the rejected
observations with $\tau<\tau_\text{min}$ generally have short exposure
times and therefore bad counting statistics, resulting in a scattered
rms-flux relation. In the following discussions of the behavior of $k$ and
$C$ we will not include these observations (but see
Sect.~\ref{subsec:hard_soft_failed}).

Even though the $\chi^2$ fits for linearity are not always so good, the
maximum deviations from the best fit straight line are generally less than
15\%. We therefore think that it is acceptable to test linearity by $\tau$
and use the fit parameters $k$ and $C$ for correlations with other
variables, although these correlations may contain some approximate only
data points due to fitting a model which does not exactly describe the
data.

\begin{figure*}
\centering
   \includegraphics[width=17cm]{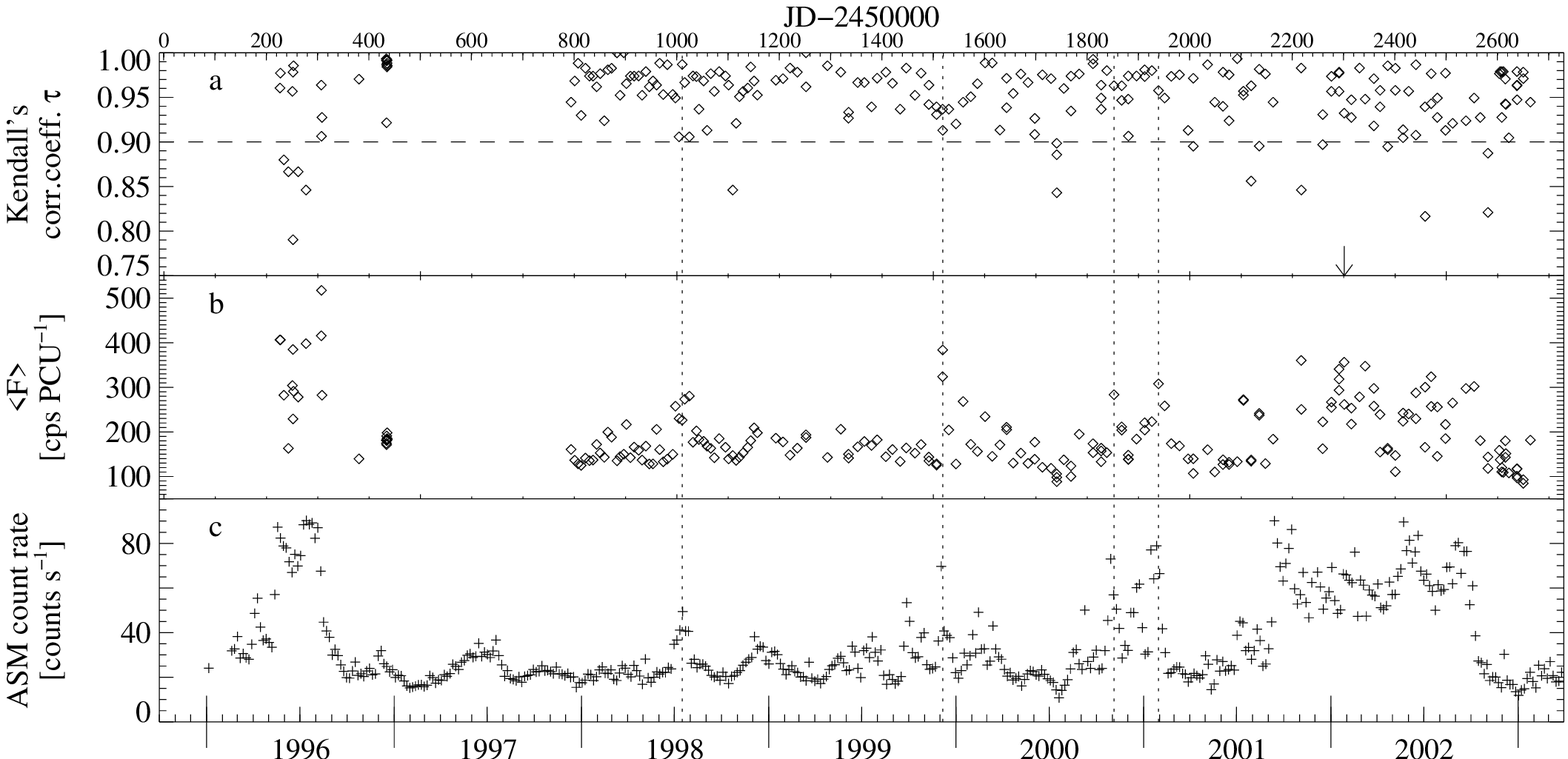}
   \caption{\textbf{a)} Kendall's $\tau$, the long-dashed line
     indicates $\tau_\text{min}$, observations with $\tau<\tau_\text{min}$
     were not included in the further analysis.  \textbf{b)} Mean
     \textsl{RXTE} \textsl{PCA} count rate for energy band 2
     ($\sim$4--6\,keV). \textbf{c)} Mean \textsl{RXTE} ASM count rate,
     rebinned to a resolution of 5.6\,days, the orbital period of Cyg~X-1
     \citep{brocksopp:99}. The four most prominent intermediate states are
     indicated by vertical dotted lines.}
   \label{fig:h4744f2}


   \includegraphics[width=17cm]{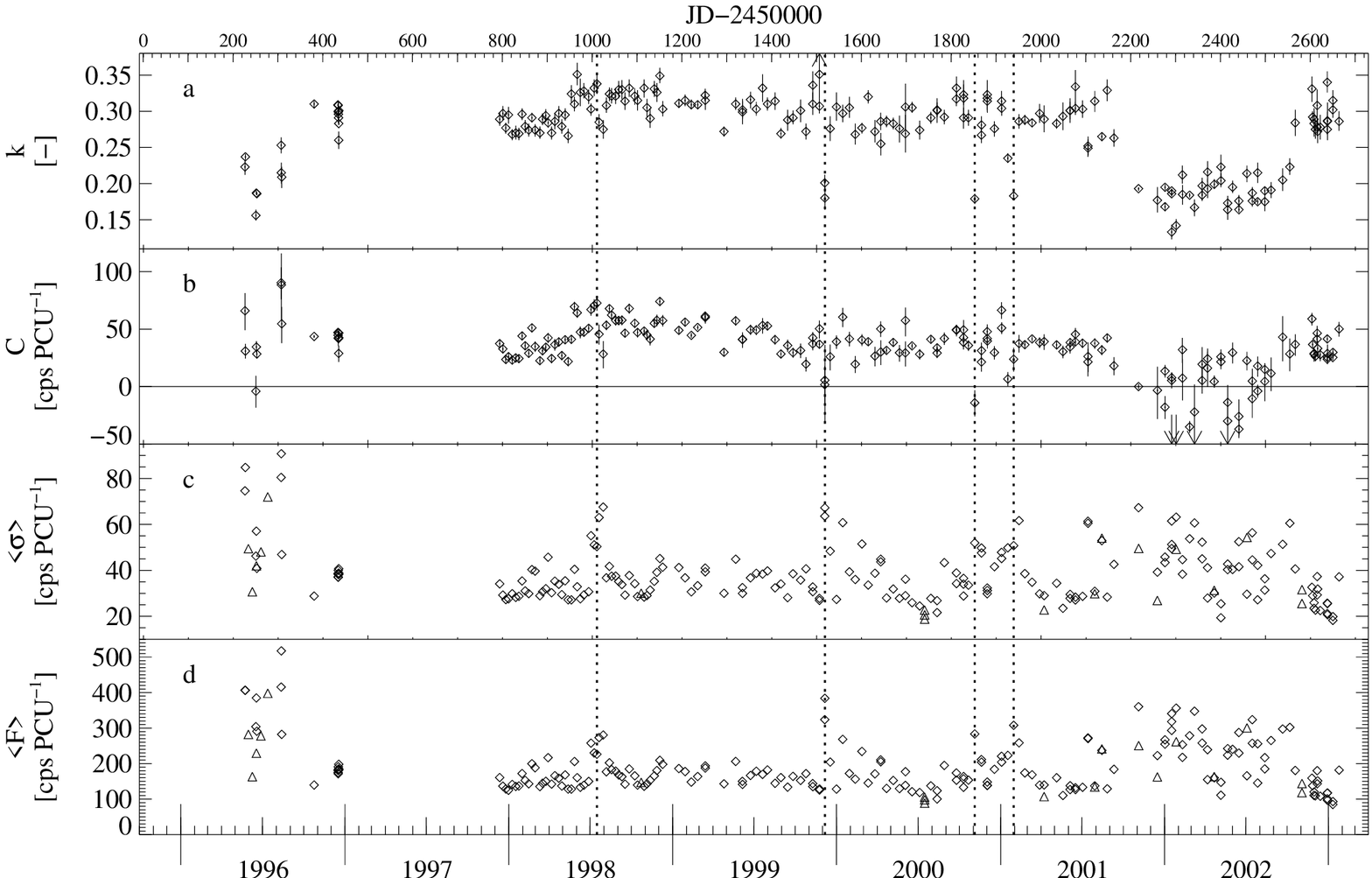}
   \caption{Fit parameters and Cyg~X-1 system properties
     for energy band 2 ($\sim$4--6\,keV). \textbf{a)} Slope, and \textbf{b)}
     $x$-axis intercept of the rms-flux-relationship, \textbf{c)} mean rms
     value, $\langle\sigma\rangle$, \textbf{d)} mean \textsl{RXTE}
     \textsl{PCA} count rate for energy band~2. Error bars given for slope
     and intercept are uncertainties at the 90\% confidence level for two
     interesting parameters. Triangles indicate observations which cannot
     be described by a linear rms-flux relation (i.e.,
     $\tau<\tau_\text{min}$).}
   \label{fig:h4744f3}

\end{figure*}

\section{Results}\label{sec:results}

\subsection{Evolution of slope and intercept}\label{subsec:evolution}
Figs.~\ref{fig:h4744f2}a--c show Kendall's $\tau$, the mean
\textsl{RXTE} \textsl{PCA} count rate, $\langle F\rangle$, for energy band 2
($\sim$4--6\,keV), and the \textsl{RXTE} All Sky Monitor (ASM;
\citealt{levine:96}) count rate for the time from 1996 to 2003.  In
Figs.~\ref{fig:h4744f2}b and c the soft state that took place 1996
March--September and the ``soft state'' of 2001 October--2002 October are
reflected by increased count rates. The fact that the failed transitions
and the soft state behavior of 2001/2002 look similar in $C$ together with
the observation that the radio emission is not always quenched in the
2001/2002 soft state (G.G.~Pooley, private communication), allows one to
speculate that the 2001/2002 soft state shows intermediate state phases
\citep[see, e.g.,][]{zdziarski:02}.  Given the uncertain nature of the
2001/2002 soft state we must be careful not to assume that it is a canonical
soft state or equal to the 1996 soft state without further proof (Wilms et
al. 2003, in prep.). Between the 1996 soft state and the 2001/2002
interval, Cyg~X-1 was usually found in the hard state, infrequently
interrupted by the ``failed state transitions'' described in the
introduction and paper~I (see also Sect.~\ref{subsec:hard_soft_failed}).

\begin{figure}
   \resizebox{\hsize}{!}{\includegraphics{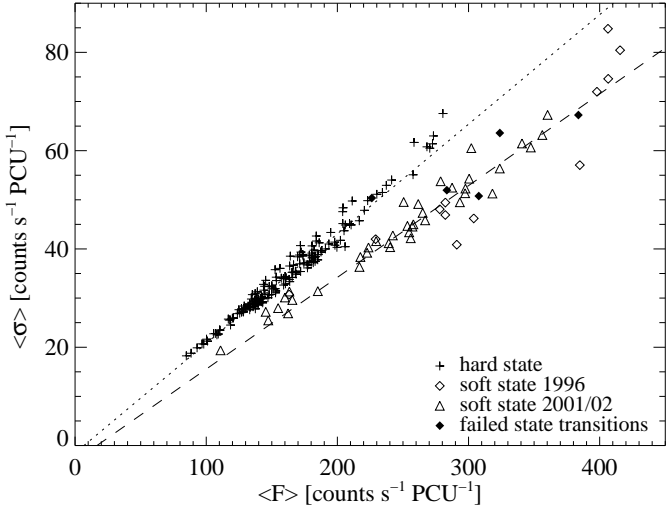}}
   \caption{Mean rms variability, $\langle \sigma\rangle$, as a
     function of mean \textsl{PCA} count rate, $\langle F\rangle$. The hard
     state observations line up to follow a linear rms-flux relationship.
     There is also a linear rms-flux relationship for the soft state
     observations, but with a slightly smaller slope.}
   \label{fig:h4744f4}
\end{figure}

Figs.~\ref{fig:h4744f3}a--d show the fit parameters $k$, $C$, mean
absolute rms variability, $\langle\sigma\rangle$, and mean \textsl{RXTE}
\textsl{PCA} count rate, $\langle F\rangle$. The most apparent correlation
is between $\langle \sigma\rangle$ and $\langle F\rangle$. To examine this
correlation, we display the rms-flux relation of the long term variations
in $\sigma$ in Fig.~\ref{fig:h4744f4}. A tight linear correlation
between $\langle \sigma\rangle$ and $\langle F\rangle$ can be seen in the
hard state and, with a slightly lower slope, in the soft state data. It
seems that the linear rms-flux relation observed on short time scales also
applies on much longer time scales. A least $\chi^2$ fit ($\chi^2=15589$
for 161 dof) of a linear model to the hard state provides
$k=0.2230\pm0.0003$ and an intercept of
$C=6.8\pm0.2\,\text{counts}\,\text{s}^{-1}\,\text{PCU}^{-1}$. These values
of $k$ and $C$ are lower than the values typically observed in the
short term rms-flux relation. The corresponding fit for the soft state data
($\chi^2=30660$ for 50 dof) provides a slope of $k=0.1867\pm0.0004$ and an
intercept at the flux axis of
$C=16.8\pm0.6\,\text{counts}\,\text{s}^{-1}\,\text{PCU}^{-1}$. The values
for the failed state transitions are associated with either the hard state
or the soft state data.

We next consider the long term behavior of $k$
(Fig.~\ref{fig:h4744f3}a). During the soft states of 1996 and
2001/2002, $k$ takes a significantly lower value than during the hard
state. During the last three ``failed state transitions'' before the
2001/2002 soft state $k$ decreases to the soft state level. The values of
the intercept show different levels, too: while in the 1996 soft state $C$
is larger than or equal to the hard state, $C$ in the 2001/2002 ``soft
state'' is smaller than in the hard state. This may be connected to the
special soft state character of the 2001/2002 interval. The detailed
characteristics of the rms-flux relation in the different states will be
discussed in the following sections.

One of the main results of paper~I was that in 1998~May a change of the
general long term behavior of Cyg~X-1 from a ``quiet hard state'' to a
``flaring hard state'' took place that coincided with a change in the PSD
shape, resulting in a \emph{decrease} of the relative rms amplitude from an
average of $36\pm1$\% to $29\pm1$\%, for the total 2--13.1\,keV power
spectrum (see paper~I, Fig.~3d). This change is seen in
Fig.~\ref{fig:h4744f3} as a rise of $k$ and $C$, while the mean rms
value, $\langle\sigma\rangle$, stays on a constant level. This apparent
discrepancy can be explained by taking into account that the total rms is
measured from $\sim 2\times 10^{-3}$\,Hz to 32\,Hz in paper~I, i.e., over a
much broader frequency range than here. As we show below
(Sect.~\ref{sec:longterm}), the drop in the global rms in 1998~May is due
to a change of the PSD outside of the frequency range covered by our
measurements of the rms-flux relationship.

\begin{figure}
  \resizebox{\hsize}{!}{\includegraphics{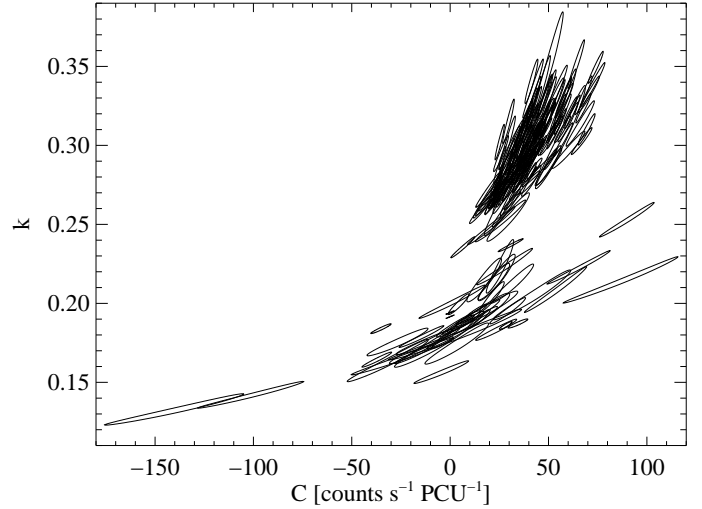}}
\caption{Contour plot at the 90\% confidence level showing the 
  two fit parameters slope and intercept at the flux axis.}
\label{fig:h4744f5}
\end{figure}

To illustrate the correlation between $k$ and $C$, Fig.~\ref{fig:h4744f5}
shows the 90\% confidence contours for these parameters. The points in the
$k$-$C$-plane fall into two separate regions corresponding to hard and soft
states.  A possible correlation between $k$ and $C$ can be seen within the
soft state, but a much stronger correlation between $k$ and $C$ is present
within the hard state.  Interestingly, the $k$-$C$ values in the hard state
fall into a distinctive, fan-like shape, which suggests that a more
fundamental relation may underly the $k$-$C$ correlation in that state.
The ``fan'' appears to converge at the values of $k$ and $C$ measured for
the long term rms-flux relation of the hard state
(Fig.~\ref{fig:h4744f4}), suggesting that the pattern of $k$ and $C$
observed in the hard state may be related to the long term rms behavior.

\begin{figure}
   \resizebox{\hsize}{!}{\includegraphics{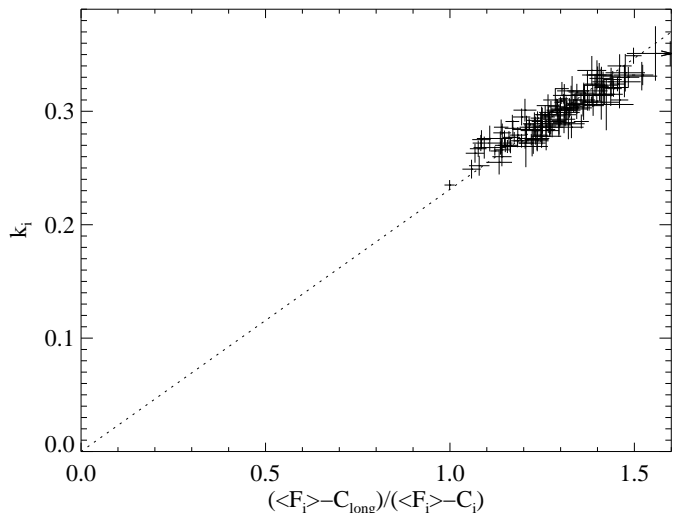}}
   \caption{For hard state observations, $k_{i}$ versus
     $(\langle F_{i}\rangle-C_\text{long})/\left(\langle
       F_{i}\rangle -C_{i}\right)$ gives a linear relation $y=bx$ with
     slope $b=0.2312$. The error bars on $k$ are the formal
     1$\sigma$ error, the horizontal error bars are calculated from
     propagating formal $1\sigma$ errors of $F_{i}$ and $C_{i}$.}
   \label{fig:h4744f6}
\end{figure}

In fact it is easy to see how a linear long term rms-flux relation can
produce the observed fan-like shape of the $k$-$C$ correlation in the
low state. First, consider that the mean $\sigma$ of the $i$th
observation, $\langle\sigma_{i}\rangle$, is related to the mean flux,
$\langle F_{i}\rangle$, by
\begin{equation}\label{eq:sigmai_short}
\langle\sigma_{i}\rangle =k_{i}(\langle F_{i}\rangle -C_{i})
\end{equation}
where $k_{i}$ and $C_{i}$ are the $k$ and $C$ values determined from the
rms-flux relation of that observation.  According to
Fig.~\ref{fig:h4744f4}, $\langle\sigma_{i}\rangle$ is also given by
\begin{equation}\label{eq:sigmai_long}
\langle\sigma_{i}\rangle =k_\text{long}(\langle F_{i}\rangle -C_\text{long})
\end{equation}
where $k_\text{long}$ and $C_\text{long}$ are the $k$ and $C$ values of the
long term rms-flux relation ($k_\text{long}=0.2230$, $C_\text{long}=6.8$).
From Eqs.~\eqref{eq:sigmai_short} and~\eqref{eq:sigmai_long} it is easy to
see that
\begin{align}\label{eq:k_rmsfluxplane}
  k_i&=\frac{k_\text{long}\left(\langle
      F_{i}\rangle-C_\text{long}\right)}{\langle
    F_{i}\rangle-C_{i}} \\
  \intertext{and}
\label{eq:c_rmsfluxplane}
C_{i}&=\frac{k_{i}\langle
  F_i\rangle-k_\text{long}\left(\langle F_{i}\rangle-C_{\rm
      long}\right)}{k_i}
\end{align}
The distribution of the $k_i$, $C_i$ values is thus caused by the
requirement to maintain the rms-flux correlation on all time scales.  The
``fan like'' distribution of the individual values is the result of the
requirement that for $k_i\rightarrow k_\text{long}$, $C_i\rightarrow
C_\text{long}$. Confirming earlier results, we find that $k_i$ is not
correlated with flux. In order to maintain the long term rms-flux
correlation, this results in a corresponding scatter of the values of
$C_i$, forming the ``fan''. We can test this directly by plotting $k_{i}$
versus $(\langle F_i\rangle -C_\text{long})/\left(\langle
  F_{i}\rangle-C_{i}\right)$ (Fig.~\ref{fig:h4744f6}), which gives a
linear relation ($\chi^2=308$ for 153 dof) with slope $0.2312\pm0.0004$
($\simeq k_\text{long}$), as expected from Eq.~\eqref{eq:k_rmsfluxplane}.
We conclude that the various values of $k$, $C$ and mean flux $\langle
F\rangle $ map out a ``fundamental plane'' in the low state, governed by
the requirement that the long term linear rms-flux relation is maintained.
We will examine the consequences of this fact in detail in
Sect.~\ref{sec:disc}.

Now that we have described the general evolution of the parameters of the
rms-flux relationship, we turn to describing the individual relations in
greater detail.

\subsection{The rms-flux relation in the hard and soft state and
  during ``failed state transitions''}\label{subsec:hard_soft_failed}
\begin{figure*}
    \centering 
  \includegraphics[width=17cm]{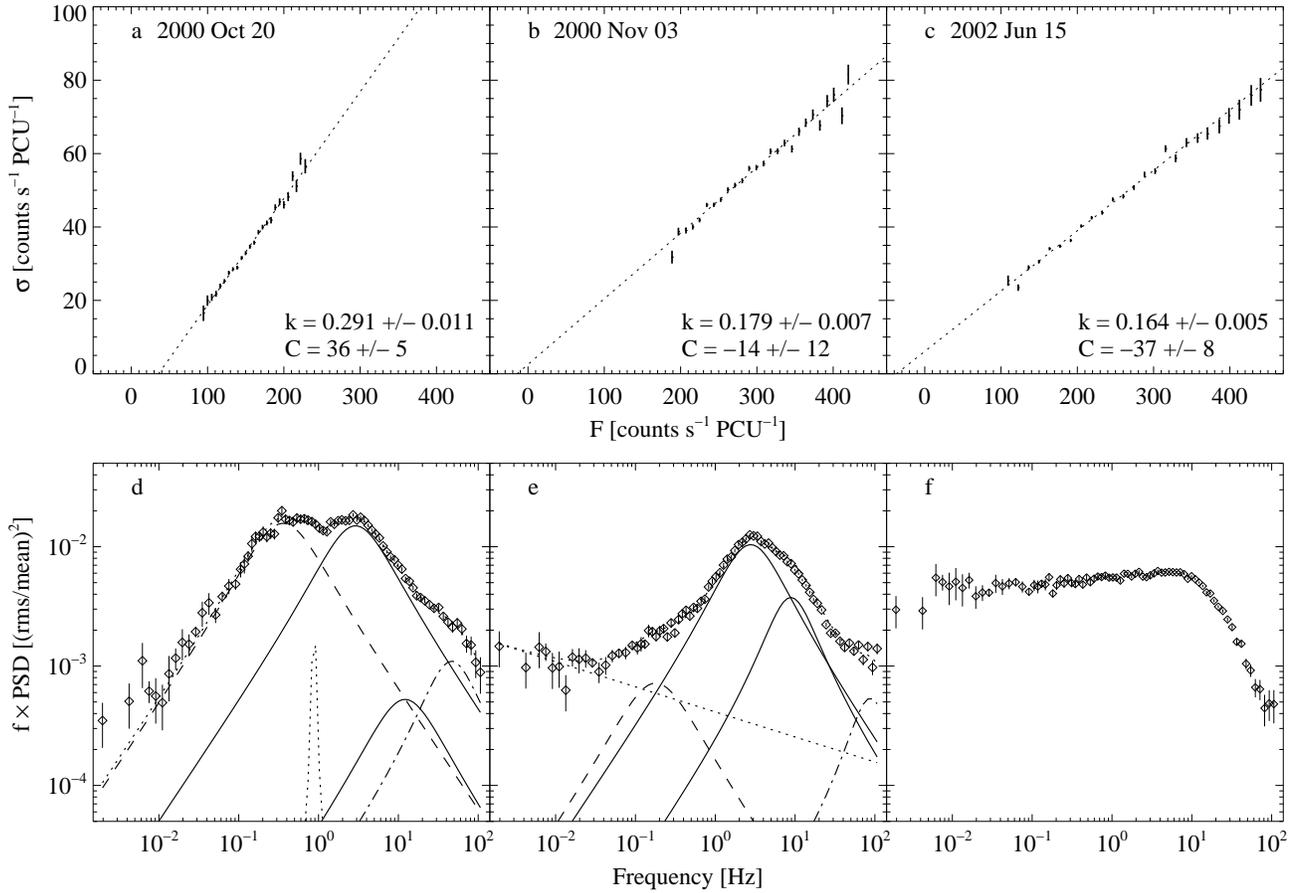}
  \caption{Linear rms-flux relations during exemplary \textbf{a)} hard
    state, \textbf{b)} ``failed state transition'', and \textbf{c)} soft
    state observations. \textbf{d)}--\textbf{f)} give the corresponding PSDs.
    Error bars of the rms variability are at the 1$\sigma$ level. See
    paper~I for a description of the PSD decomposition into single
    Lorentzians (does not work for soft state observation \textbf{f)}).}
  \label{fig:h4744f7}
\end{figure*}

The high values of Kendall's $\tau$ in Fig.~\ref{fig:h4744f2}a
prove that a linear rms-flux relation is valid throughout all states,
the hard state \emph{and} the soft state \emph{and} the ``failed state
transitions'', with relatively few outliers. Note, however, that a
number of soft state observations, predominantly during the atypical
2001/2002 ``soft state'', have rms-flux relations with a negative flux
offset $C$ (see Fig.~\ref{fig:h4744f3}b). This negative offset
cannot be explained if, for these cases, $C$ simply represents a
constant component to the lightcurve, but can easily be explained if
$C$ represents a component with \emph{constant rms}. In
Fig.~\ref{fig:h4744f7}a--c we give examples for the overall rms-flux
linearity.

\begin{figure*}
  \includegraphics[width=12cm]{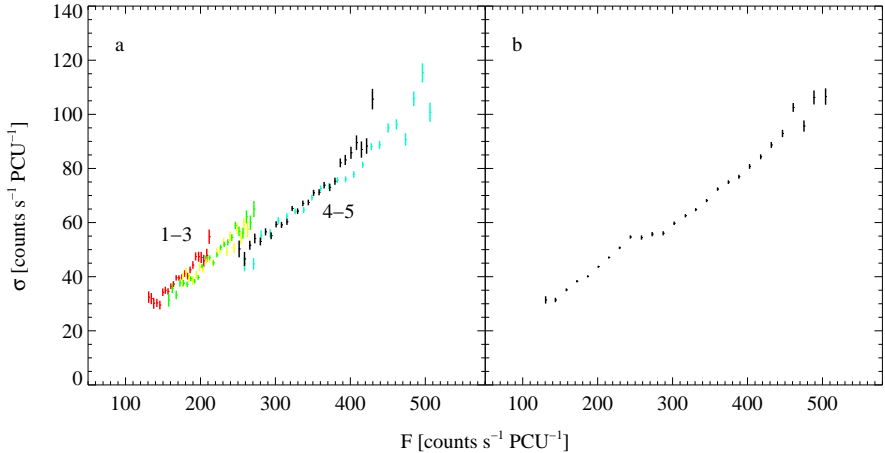}
  \caption{The observation of 2002 Feb 25 consists of five short segments
    1--5. \textbf{a)} Segments 1--3 are at lower flux, while 4 and 5
    are at higher flux, so that the rms-flux relations for each part
    are shifted with respect to each other.  \textbf{b)} The
    superposition of all parts results in a ``wavy'' rms-flux
    relation.  Error bars are at the 1$\sigma$ level.}
  \label{fig:h4744f8}
\end{figure*}

There are a small number of observations which cannot be described
with a linear model. These deviations from a linear relation have two
principal reasons: (1) As discussed in Sect.~\ref{subsec:rmsvsflux},
many observations with $\tau<\tau_\text{min}$ simply have a short
observation time, so that the number of segments per bin is relatively
low which produces the strong scatter, resulting in a poor
correlation.  (2) In contrast, the ``wavy'' rms-flux relation shown in
Fig.~\ref{fig:h4744f8}b results from short term variations in the
parameters $C$ and/or $k$ of a linear rms-flux relation.  If a single
observation (on time scales of a few 100\,s) consists of several
parts, each with different average flux (see Fig.~\ref{fig:h4744f8}a),
then the superposition of the linear rms-flux relations of all parts
will result in a ``wavy'' rms-flux relation, as each part displays a
slightly different $C$ and/or $k$ value.

To describe the flaring events when Cyg~X-1 attempts the change from
the common hard state into the soft state, without reaching the soft
state, we coined the term ``failed state transitions''
(\citealt{pottschmidt:00a}; paper~I).  Depending on how far its state
evolves, during these failed transitions Cyg~X-1 can reach the
intermediate state as defined by \citet{belloni:96}, or its behavior
can remain hard state-like during these flares.

We examined four ``failed state transitions'' that were covered by
\textsl{RXTE} observations in detail: 1998 July 15, 1999 December 05
(extracted as two separate lightcurve parts), 2000 November 03, and 2001
January 29 \citep[probably a short soft state, see][]{cui:02a}. The values
of $k$ and $C$ of the first failed state transition on 1998 July 15 are not
different from the surrounding hard state observations. During the latter
three failed state transitions, however, the values of $k$ and $C$ change
to the soft state level. For one exemplary observation on 2000 November~03,
this behavior is displayed in Fig.~\ref{fig:h4744f7}b.  Comparing the
failed state transition with the hard state observation immediately before
(Fig.~\ref{fig:h4744f7}a), the rms-flux relation of 2000 November~03
changes to a lower slope and a small, \emph{negative} intercept $C$ during
the ``failed state transition'' itself.  The PSD of this observation shows
that this change in the rms-flux relation is accompanied by a decreasing
Lorentzian component $L_1$ and an increasing Lorentzian component $L_{3}$
(Fig.~\ref{fig:h4744f7}d--e).

\subsection{Spectral dependence of the rms-flux relation}\label{subsec:spectral} 
\begin{figure}
   \resizebox{\hsize}{!}{\includegraphics{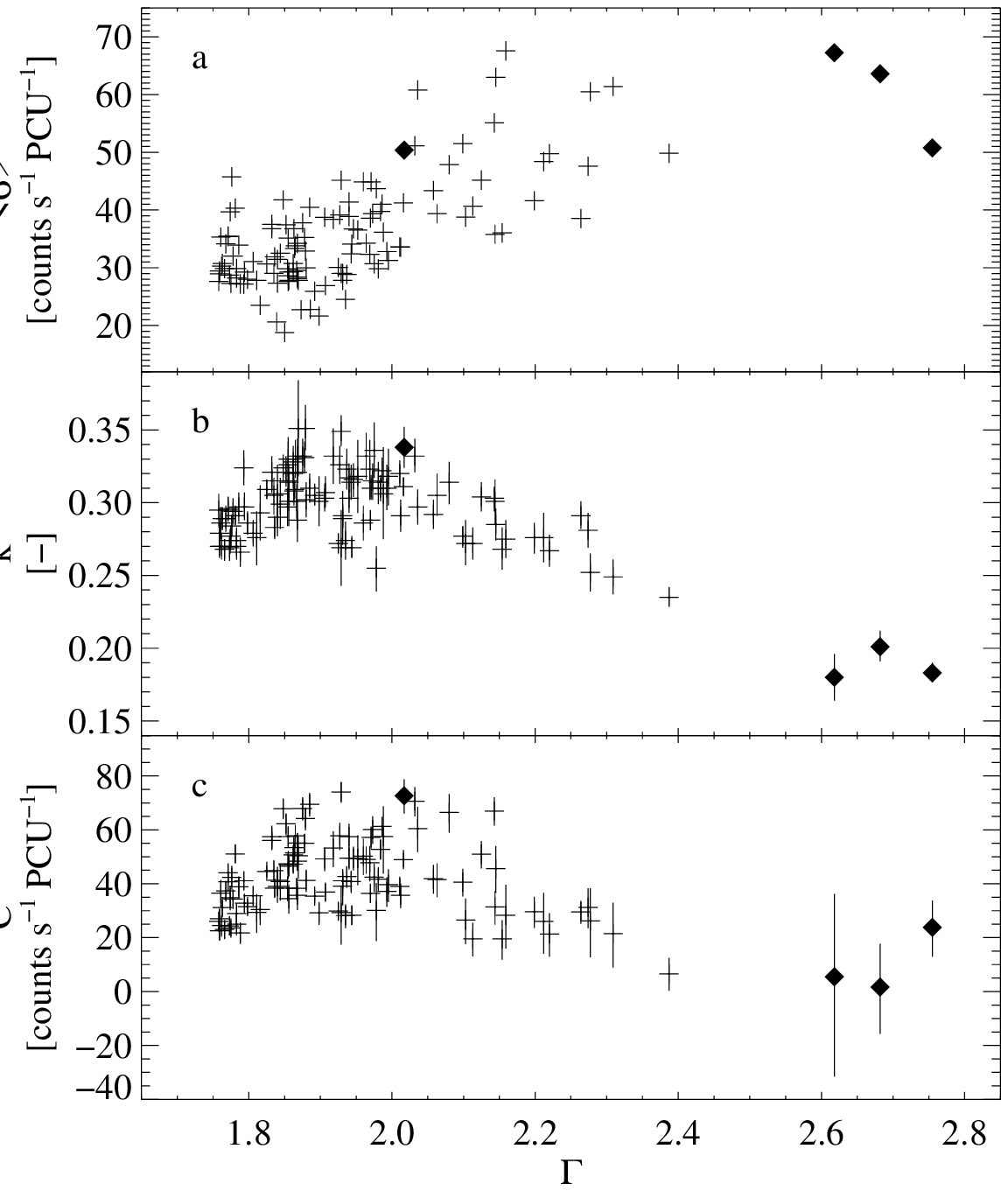}}
   \caption{\textbf{a)} Mean rms variability, $\langle\sigma\rangle$,
     \textbf{b)} slope, $k$, and \textbf{c)} flux intercept, $C$, as a
     function of photon spectral index, $\Gamma$, for failed state
     transitions and the hard state observations. Diamonds designate
     ``failed state transitions'', crosses are for hard state observations.
     Error bars for $k$ and $C$ are given at the 90\% confidence level for
     two interesting parameters.}
   \label{fig:h4744f9}
\end{figure}

We confirm a correlation between the photon spectral index, $\Gamma$, i.e.,
the hardness of the spectrum, and the mean rms, $\langle \sigma \rangle$,
of the examined observations. In the hard state, the photon spectrum of
each observation can roughly be described as the sum of a power law
spectrum $E^{-\Gamma}$ with photon index $\Gamma$ and a multi-temperature
disk black body \citep{mitsuda:84a}.  The correlation of
Fig.~\ref{fig:h4744f9}a takes into account only hard state and ``failed state
transition'' observations and shows that $\langle\sigma\rangle$ increases
as the spectrum grows softer. As rms variability and flux are linearly
correlated in the hard state, this correlation between $\Gamma$ and
$\langle\sigma\rangle$ is equivalent to a correlation between $\Gamma$ and
flux.  A similar result with \textsl{RXTE} ASM data of Cyg~X-1 has been
recently shown by \citet{zdziarski:02}, who found a very strong
hardness-flux anticorrelation in the hard state.

We note that we are using the definition of paper~I, using the term
``failed state transitions'' for those observations which exhibit a
significantly increased time lag compared to its typical value, since the
X-ray time lag seems to be a more sensitive indicator for state changes
than the X-ray spectrum. As has been already been noted in
Sect.~\ref{subsec:hard_soft_failed} there are only four corresponding
observations.

The filled diamond at $\Gamma\sim 2$ in Fig.~\ref{fig:h4744f9} belongs to the
``failed state transition'' of 1998 July~15. The fact that this data point
is situated within the value range of hard state observations in
Figs.~\ref{fig:h4744f4} and~\ref{fig:h4744f9} confirms that this ``failed
state transition'' has to be treated separately and classified as being
close to the hard state, as already mentioned in
Sect.~\ref{subsec:hard_soft_failed}.  Allowing for the split in values of
$k$ and $C$, respectively, between the hard state and failed state
transition data, any correlation between the photon index and $k$, or $C$,
cannot easily be detected.

\begin{figure}
   \resizebox{\hsize}{!}{\includegraphics{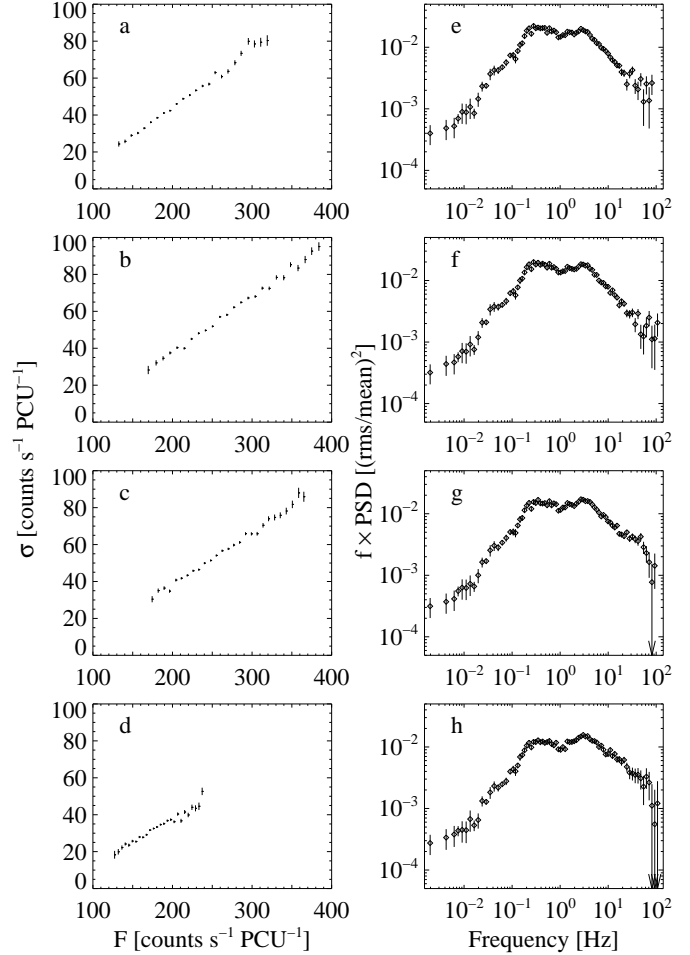}}
   \caption{Shape of \textbf{a)}-\textbf{d)} the rms-flux relation and
     \textbf{e)}-\textbf{h)} the corresponding PSD for the observation of 1999
     May~21 for energy bands 2--5, respectively.}
   \label{fig:h4744f10}
\end{figure}

So far we concentrated on the rms-flux relation in the second lowest energy
band, i.e., energy band 2 in the range $\sim$4--6\,keV. This approach is
justified as the linear rms-flux relation is found to be valid for all
energy bands considered here (see, e.g., Fig.~\ref{fig:h4744f10}), such that
the general behavior described for energy band~2 is also seen in the other
bands. Even the rms-flux relation in energy band~3 which contains the iron
K line does not show a different behavior, in agreement with the results of
\citet{maccarone:02a} that the iron line tracks the continuum at least in
the soft and transition states. Nevertheless, there are subtle differences
in the shape of the rms-flux correlation.

\begin{figure*}
\centering
   \includegraphics[width=17cm]{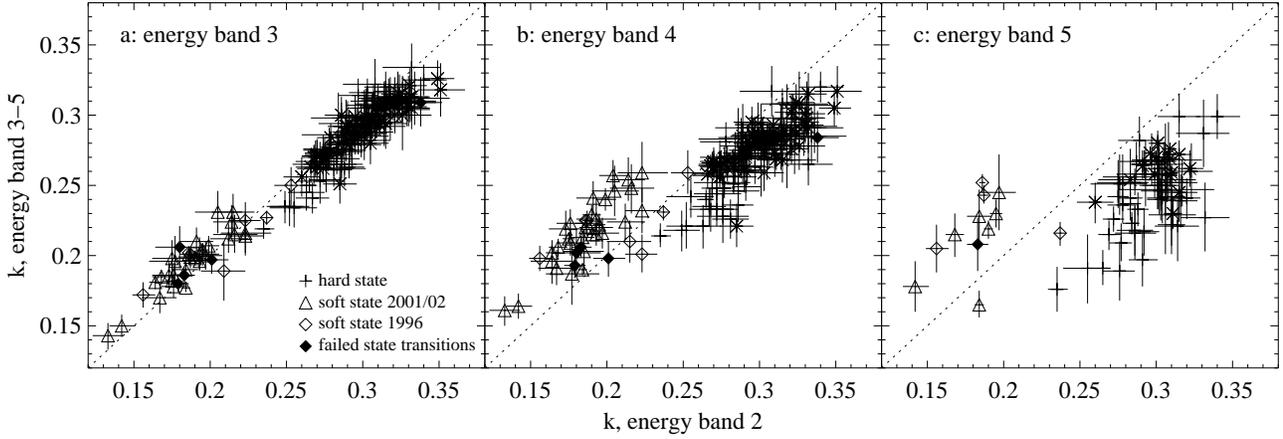}
   \caption{Slope of energy bands 3--5 plotted against the slope of
     energy band~2. The dotted line indicates $k_i=k_2$ where
     $i=\text{band~3}$ through~5. Error bars are at the 90\% confidence
     level for two interesting parameters.}
   \label{fig:h4744f11}
\end{figure*}

\begin{figure*}
  \centering \includegraphics[width=17cm]{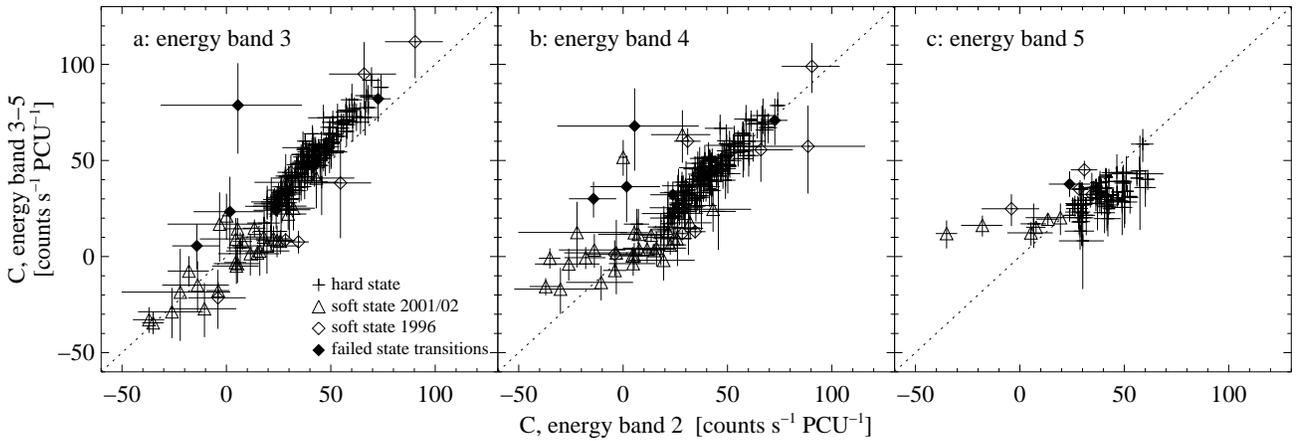} \caption{Behavior of $C$
    for energy bands~3 through~5, as a function of $C$ measured for energy
    band~2. The dotted line indicates $C_i=C_2$ where $i=\text{band~3}$
    through~5. Error bars are at the 90\% confidence level for two
    interesting parameters.}  \label{fig:h4744f12}
\end{figure*}

Comparing the values of $k$ and $C$ for the different energy bands, our
energy resolved results for the rms-flux correlation show that $k$, which
represents the fractional rms variability of the time variable component of
the emission in the 1--32\,Hz band, generally decreases with energy when
the source is in the hard state (Fig.~\ref{fig:h4744f11}), in agreement with
earlier results \citep[e.g.,][]{nowak:98a}. The same is true for $C$ (see
Fig.~\ref{fig:h4744f12}). For the highest energy band~5, our analysis reveals that $k$ and $C$ only
slightly correlate with energy band~2 (Figs.~\ref{fig:h4744f11}c
and~\ref{fig:h4744f12}c). This result confirms earlier work claiming only a
`loose coupling' between the source variability at low and at high
energies \citep{zycki:03a,churazov:01a,maccarone:00a,gilfanov:99a}. In
terms of Compton corona models, the fraction of photons from the accretion
disk significantly decreases when proceeding from energy band~2 to energy
band~5. Therefore we take our result to confirm earlier claims that the
variability properties of Cyg~X-1 are driven by coronal fluctuations and
not by changes in the soft photon input to the putative Compton corona
\citep[and references therein]{maccarone:00a,churazov:01a}.

\begin{figure}
\resizebox{\hsize}{!}{\includegraphics{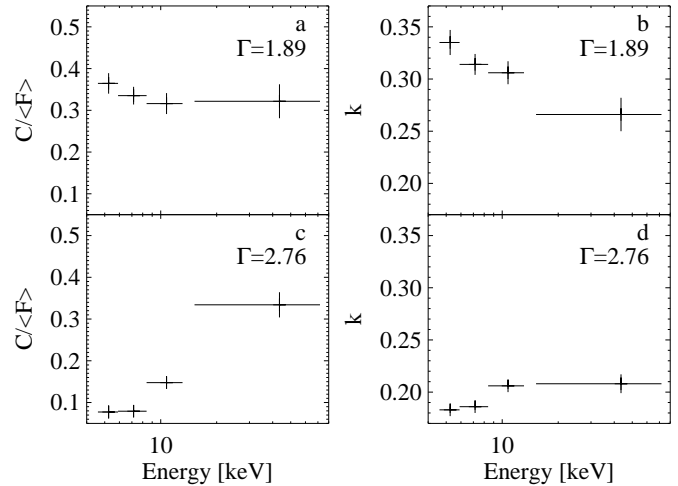}}
\caption{Spectral shapes of the two variability components characterized by
  $C$ and $k$, respectively. $C$ has been normalised by $\langle F\rangle$.
  \textbf{a)}-\textbf{b)} show a typical hard state observation (2001 April
  7) with $\Gamma=1.89$, whereas \textbf{c)}-\textbf{d)} give an example of
  the tilted spectral shapes which can be found for higher $\Gamma$
  (``failed state transition'' observation of 2001 Jan 29 with
  $\Gamma=2.76$).}
\label{fig:h4744f13}
\end{figure}

To examine the overall spectral shapes of the two variability components
characterized by $C$ and $k$, respectively, we plotted these values for the
energy bands 2--5, normalising $C$ by $\langle F\rangle$ to account for the
instrument response. Fig.~\ref{fig:h4744f13} shows two examples with
different photon spectral index, $\Gamma$. For most observations with
$\Gamma$ lower than $\sim 2.0$, which means for the majority of hard state
observations, the spectral shapes of $C$ and of $k$ are similar, but the
spectral shape of $C$ tends to be somewhat flatter: both $C$ and $k$ show a
soft energy spectrum, i.e., higher values at low energies and lower values
at higher energies (Figs.~\ref{fig:h4744f13}a--b). When we proceed to
higher $\Gamma$, observations begin to display a flat or even tilted
spectrum, i.e. lower values at low energies and higher values at higher
energies, more pronounced for $C$ than for $k$
(Figs.~\ref{fig:h4744f13}c--d). The change in the shape of the
normalised $C$ (cf.  Figs.~\ref{fig:h4744f13}a and c) is caused by the
combination of an intrinsic change in the spectral shape of the
unnormalised $C$ {\it and} a change in the spectral shape of
$\langle F \rangle$.  Suggestions that the two variability components $C$ and
$k$, representing a constant and a variable rms component, will have
basically different energy shapes cannot be confirmed, but to a certain
degree in the low state the spectral shape of $C$ is flatter and harder
than the spectral shape of $k$.

\subsection{The rms-flux relation on short time scales} 
\begin{figure}
   \resizebox{\hsize}{!}{\includegraphics{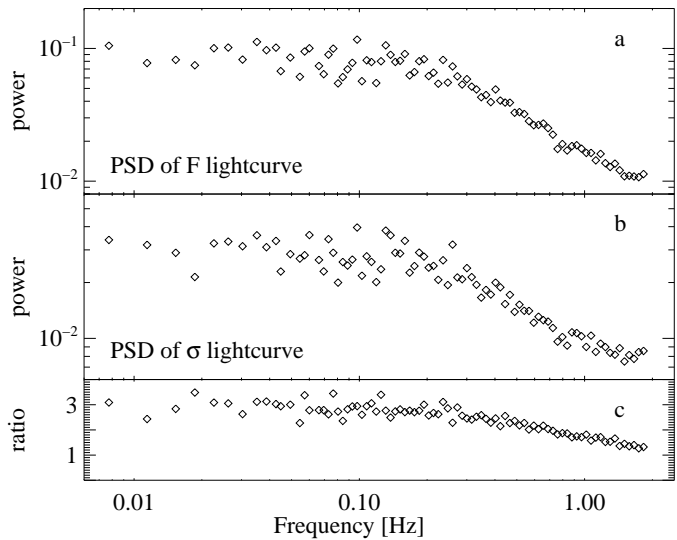}}
   \caption{PSDs of \textbf{a)} $F$ lightcurve and \textbf{b)} $\sigma$
     lightcurve for the hard state observation on 1999 July 05. The
     original flux lightcurve with time resolution of $\Delta t=2^{-8}$\,s
     has been divided into 0.25\,s segments, and the frequency range
     $\nu=4$--128\,Hz was chosen for calculation of the rms variability of
     each segment. To compute PSDs from the flux and the rms lightcurves --
     consequently both with the time resolution of the segment length
     $\Delta t=0.25$\,s -- segments of 1024 time elements ($=1024 \times
     0.25\,\text{s}=256$\,s) were used, thus resulting in a frequency range
     from $\nu_\mathrm{min}=0.004$\,Hz to $\nu_\mathrm{max}=2$\,Hz. We
     rebinned this range to 84 logarithmically spaced bins. \textbf{c)}
     Ratio of the $F$ lightcurve PSD to the $\sigma$ lightcurve PSD in
     order to quantify the similarity of the two PSDs.}
   \label{fig:h4744f14}
\end{figure}

The fact that $\sigma$ tracks the flux implies that the ``lightcurve'' of
the rms values is similar to the flux lightcurve. This fact can be used to
probe how the rms responds to flux changes on short time scales (see also
Uttley et al. 2003, in prep.). For example, if the rms does indeed track
variations on all time scales, the PSD of the ``rms lightcurve'' should
look similar to the PSD of the conventional ``flux lightcurve''.  In
contrast, if the rms only tracks variations on time scales longer than,
say, 10\,s, the PSD of the rms lightcurve should be sharply cut off at
0.1\,Hz.  In order to test this issue we created 0.25\,s lightcurve
segments and calculated $F$ and $\sigma$ for each of these segments to make
flux and rms lightcurves.  From these lightcurves, we calculated the PSDs
(shown for an arbitrarily chosen observation in Fig.~\ref{fig:h4744f14}). In
this case, we do not correct the rms lightcurve for Poisson noise, since
the stochastic nature of the noise process means that such a correction is
only possible when averaging over comparably long time intervals.  Note,
however, that the effect of the Poisson noise and intrinsic noise
contributions to the rms lightcurve is to add power to the rms PSD at high
frequencies, not to cause spurious high-frequency cut-offs or breaks.

Over the frequency range $\sim$0.01--0.2\,Hz the rebinned PSDs show the
same features as indicated by the flat ratio in Fig.~\ref{fig:h4744f14}c.
Therefore it is likely that the rms-flux relation is fulfilled on
time scales from 5\,s to 100\,s.  Above 0.2\,Hz, however, Poisson noise
starts to dominate the rms PSD such that the shape of the rms and flux PSDs
cannot be compared.

Similar tests dealing with the variability on different time scales
have been discussed by \citet{mineshige:94a} and \citet{maccarone:02b}.

\section{Discussion}\label{sec:disc}

\subsection{The origin of the rms-flux relation}
\citet{uttley:01} have interpreted the rms-flux relation in
terms of the disk fluctuation model of \citet{lyubarskii:97a}, where
variations in the accretion rate occur at various radii and propagate
inwards to modulate the X-ray emission.  The fact that variations at
larger radii should have longer characteristic time scales naturally
explains the fact that the short term rms variations are modulated by
longer time scale flux changes.  As the slower variations in the
accretion flow propagate inwards they are able to modulate the
variations on shorter time scales. \citet{kotov:01a} have extended this
model by introducing an extended X-ray emitting region with a
temperature gradient, to explain the frequency dependent time lags
between hard and soft bands observed in Cyg~X-1
\citep{pottschmidt:00a}.

\citet{churazov:01a} have pointed out that in a standard thin disk,
viscous damping will smooth accretion rate fluctuations on a viscous
time scale, i.e., before they reach the inner disk, implying that for
the \citet{lyubarskii:97a} model to work, the fluctuations in
accretion rate must happen in an optically thin or geometrically thick
flow, such as an advection dominated accretion flow (ADAF) or another
type of thick disk.  We have shown that the linear rms-flux relation
is seen in both high and low flux states, so that it very well might
be associated with some component of the system that is common to both
states.  Since the X-ray variability during the soft state appears to
be associated with the power law emission rather than thermal emission
from the disk \citep[which may be constant,][]{churazov:01a}, it seems
likely that the rms-flux relation is associated directly with a corona
which is present in both hard and soft states.  We therefore speculate
that the rms-flux relation in Cyg~X-1 (and probably also other objects
showing linear rms-flux relations) may present evidence for accretion
rate variations in a \emph{fluctuating coronal accretion flow} \citep[see,
e.g.,][and references therein]{smith:01b, klis:01a}. The
suggestion that the corona itself accretes has been used to explain a
number of observational features of both AGN and galactic BH systems
\citep{witt:97}, and since an optically thin coronal accretion flow is
similar to an ADAF, such a model would not encounter the problems
faced by a thin disk origin for the accretion variations
\citep{churazov:01a}.

\subsection{The long term rms-flux relation}\label{sec:longterm}
In the previous sections we have shown that the linear rms-flux
relation seems to apply on very long time scales, since the average
1--32\,Hz rms of an observation scales linearly with the average flux
of the observation.  The hard and soft state data form two separate
long term rms-flux relations, with different slopes $k_\text{long}$
(somewhat lower in the soft state, $k_\text{long}=0.19$ versus
$k_\text{long}=0.22$ in the hard state).  Intriguingly, the
flux-offsets $C_\text{long}$ in the long term rms-flux relations of
both hard and soft states are very close to zero, and certainly of
much smaller amplitude than the typical offsets observed in the
short term rms-flux relations of either state.  The scatter in the
hard state long term rms-flux relation is particularly small.

Combined with the almost-zero offset of the long term relation, this
tight correlation implies that the fractional rms of the 1--32\,Hz
variability in the hard state is remarkably constant (similar to
$k_\text{long}$, i.e., $\sim$22\%).  This result is rather surprising
because the hard state PSD is clearly not stationary, i.e., the
Lorentzians which successfully describe the PSD change significantly
in both peak frequency and fractional rms (paper~I).  The most obvious
example of such a change in the PSD can be seen in 1998~May (see
paper~I, Fig.~3) when the peak frequencies of the Lorentzians
simultaneously increase, and the fractional rms of the $L_{3}$
Lorentzian (which lies between 3--10\,Hz) and the fractional rms of
the total PSD both decrease significantly.  However, despite such a
marked change in the PSD, the data before and after 1998~May lie along
the same long term hard state rms-flux relation.  The fact that the
fractional rms in the 1--32\,Hz band seems undisturbed by significant
changes in the PSD suggests some sort of conservation of fractional
rms at high frequencies, even though the low-frequency (and hence the
total) fractional rms changes significantly.  Although the specific
1--32\,Hz band chosen to measure the rms-flux relation is somewhat
arbitrary, we point out that the bulk of the power in this band is
associated with the $L_2$ and $L_3$ Lorentzians, and the band fully
encompasses the range of frequencies between 3--10\,Hz in which a
strong and well-defined anti-correlation between Lorentzian peak
frequency and fractional rms is clearly seen (see paper~I, Fig.~6).
We suggest, therefore, that some process which conserves fractional
rms above a few~Hz may be at work, with fractional rms in the $L_{3}$
Lorentzian decreasing as the $L_{2}$ Lorentzian moves further into the
1--32\,Hz band, resulting in a constant fractional rms in that band.
We now examine how such a phenomenological model relates to the
observed fundamental plane of short term rms-flux relation parameters
$k$, $C$, and flux.

\subsection{The hard state $k$-$C$-$\langle F\rangle$ fundamental plane}
We have shown that the ``fan shape'' of the correlation between the
rms-flux relation slope and its flux-offset in the hard state can be
directly related to the long term rms-flux relation.  The fact that
$k$ and $C$ vary but the average 1--32\,Hz rms \emph{must} track the
average flux explains the degeneracy in the correlation, which is the
result of the various fluxes observed for a given pair of $k$ and $C$
values.  Thus $k$, $C$, and $\langle F\rangle$ track out a fundamental
plane which describes the forms the rms-flux relation can take in the
hard state. 

As noted by \citet{uttley:01}, $C$ can be interpreted either simply as a
constant-flux component to the lightcurve, or as a constant rms component.
We concentrate on these two interpretations since they are the simplest
possibilities, but it should be emphasized that there are numerous other
forms the variability could take, e.g., a component of constant fractional
rms.

We have shown that, in the soft state at least, $C$ can be negative, which
is only possible if the offset in this state represents a constant rms
component.  We argue that the existence of the $k$-$C$-$\langle F\rangle$
fundamental plane also suggests that $C$ represents a constant rms
component in the hard state.  For example, consider an increase in the
slope $k$ between two observations, while keeping the average flux of the
observations constant.  The increase in $k$ corresponds to an increase in
the fractional rms of the linear rms-flux component of the lightcurve.  In
order to maintain the constant total fractional rms as given by the
observed long term rms-flux relation, $C$ must also increase.  Therefore we
suggest that both a constant rms component and a linear rms-flux component
exist in the lightcurve and that the relative contribution of each varies
in such a way that the sum of both components, the total fractional rms, is
constant.

As a constraint to possible models it should be mentioned, as already noted
in Sect.~\ref{subsec:spectral}, that the spectral shapes of $k$ and $C$ are
generally similar, with $C$ tending to a flatter/harder spectrum, and there
is no obvious correlation between spectral index $\Gamma$ and $C$,
$C_\text{long}$, and $k$, respectively.

\subsection{Summary}
To summarize, in this paper we analyzed Cyg~X-1 lightcurves from 1996
to 2003 in an attempt to describe the long term evolution of X-ray rms
variability.  Our main results are
\begin{enumerate}
\item The linear rms-flux relation is valid throughout all states and
  energy ranges observed.
\item The slope of the linear rms-flux relation is steeper in the hard
  state than in the soft and intermediate states.
\item The rms-flux relation is valid not only on short time scales
  (seconds) but also on longer time scales (weeks--months). 
\item In the hard state, the slope of the flux-rms-relationship is
  correlated with its flux intercept, but the correlation is
  increasingly degenerate at higher values of $k$ and $C$, forming a
  fan-like distribution in the $k$-$C$ plane.  The hidden parameter
  governing the degeneracy is the long term average flux, $\langle
  F\rangle$, and the $k$-$C$-$\langle F\rangle$ relation maps out a
  ``fundamental plane'' in the low state.
\item The intercept of the linear rms-flux relation is relatively
  constant and positive in the hard state, but is diverse in the soft
  state: in the 1996 soft state $C$ is positive and slightly higher
  than during the hard state, in the atypical 2001/2002 ``soft state''
  $C$ is decreasing relative to the hard state level, with several
  observations showing a negative $C$. During ``failed state
  transitions'', $C$ is smaller than in the hard state, similar to the
  soft state behavior.
\item The $F$-$\sigma$-$\Gamma$ correlation can be confirmed for the hard
  state.
\item The clear dichotomy of $k$ in a hard and a soft state level in
  energy band~2 is weaker at higher energies.
\end{enumerate}

\begin{acknowledgements}
  This work has been financed by grant Sta~173/25 of the Deutsche
  Forschungsgemeinschaft. We thank Sara Benlloch for helpful discussions
  and Ron Remillard for generously supplying TOO data.  This research has
  made use of data obtained from the High Energy Astrophysics Science
  Archive Research Center (HEASARC), provided by NASA's Goddard Space
  Flight Center. JW thanks the Instituto Nacional de Pesquisas Espaciais,
  S\~ao Jos\'e dos Campos, Brazil, for its hospitality during the
  finalization of this paper.
\end{acknowledgements}

\appendix

\section{Buffer overflows in \textsl{RXTE} binned data modes}
\begin{figure}
\resizebox{\hsize}{!}{\includegraphics{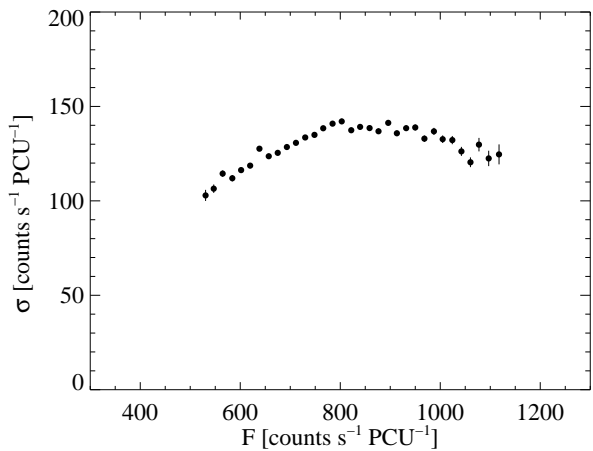}}
\caption{In energy band 1 the rms-flux relation for observation 2000
  Nov~03, just as several other high flux observations during
  intermediate and soft states, displays an ``arch-like'' form. The given
  error bars are at the 1$\sigma$ level.} \label{fig:h4744fa1}
\end{figure}

During phases of high soft X-ray flux, the rms-flux relation for
energy band~1 ($\sim$2--4.6\,keV) shows a clear deviation from its
typical linear behavior. These phases are characterized by a smaller
rms than what would be expected when extrapolating the rms-flux
relation measured for smaller fluxes (Fig.~\ref{fig:h4744fa1}). We
initially associated this breakdown of the relationship with changes
in the Comptonizing corona during phases of high luminosity
\citep{gleissner:02a}, however, this interpretation was wrong.

Instead, the ``arches'' are caused by buffer overflows in the \textsl{PCA}
hardware \citep[see also][]{gierlinski:03a,vanstraaten:03a}.  For the
binned data mode B\_2ms\_8B\_0\_35\_Q, with which most of our low
energy data are taken, 4\,bit wide counters are used to form the
binned spectrum during each 2\,msec binning interval. During phases of
very high flux, which are especially likely during short outbursts
possible during the failed state transitions, these buffers will
overflow. Due to the softness of the X-ray spectrum during the failed
state transitions, the likelihood for such overflows is highest in the
lowest energy band. The buffer overflows become apparent when
comparing the lightcurve of the binned data mode with the lightcurve
as determined, e.g., from the GoodXenon data, which does not suffer
from these problems. Except for phases of high flux, these lightcurves
agree. During phases of high flux, however, multiples of 16\,events
are missing in the lightcurve from the binned data.  Since the buffer
overflows only occur for a few bins of the high resolution lightcurve,
the flux determination is barely affected, however, the rms is reduced
and ``arches'' such as that shown in Fig.~\ref{fig:h4744fa1} are observed.
For the computation of the rms-flux relationship, we therefore
resorted to using energy band 2, where the overall photon flux is
lower and buffer overflows do not occur. 

Note that GoodXenon data are not available for all observations considered
here, furthermore, the energy resolution of the GoodXenon data is not good
enough to allow the study of other timing parameters. We note that the
rms-flux relationship is especially sensitive to buffer overflows as the
data are sorted according to flux. For the determination, e.g., of power
spectra, longer lightcurves are chosen, and the buffer overflows are only
apparent in a slight change of the rms level of the power spectrum.
Therefore, buffer overflows at the level seen in the data here only very
slightly affect the normalization of the PSDs such as those used in paper~I
and do not have any influence on our earlier results.


\begin{thebibliography}{47}
\expandafter\ifx\csname natexlab\endcsname\relax\def\natexlab#1{#1}\fi

\bibitem[{Belloni \& Hasinger(1990)}]{belloni:90b}
Belloni, T. \& Hasinger, G. 1990, A\&A, 230, 103

\bibitem[{Belloni {et~al.}(1996)Belloni, {M\'endez}, {van der Klis}, Hasinger,
  Lewin, \& {van Paradijs}}]{belloni:96}
Belloni, T., {M\'endez}, M., {van der Klis}, M., {et~al.} 1996, ApJ, 472, L107

\bibitem[{{Brocksopp} {et~al.}(1999){Brocksopp}, {Fender}, {Larionov},
  {Lyuty}, {Tarasov}, {Pooley}, {Paciesas}, \& {Roche}}]{brocksopp:99}
{Brocksopp}, C., {Fender}, R.~P., {Larionov}, V., {et~al.} 1999, MNRAS, 309,
  1063

\bibitem[{{Churazov} {et~al.}(2001){Churazov}, {Gilfanov}, \&
  {Revnivtsev}}]{churazov:01a}
{Churazov}, E., {Gilfanov}, M., \& {Revnivtsev}, M. 2001, MNRAS, 321, 759

\bibitem[{{Cui} {et~al.}(2002){Cui}, {Feng}, \& {Ertmer}}]{cui:02a}
{Cui}, W., {Feng}, Y., \& {Ertmer}, M. 2002, \apjl, 564, L77

\bibitem[{{di Matteo} \& {Psaltis}(1999)}]{dimatteo:99a}
{di Matteo}, T. \& {Psaltis}, D. 1999, ApJ, 526, L101

\bibitem[{Dove {et~al.}(1998)Dove, Wilms, Nowak, Vaughan, \&
  Begelman}]{dove:97c}
Dove, J.~B., Wilms, J., Nowak, M.~A., Vaughan, B., \& Begelman, M.~C. 1998,
  MNRAS, 289, 729

\bibitem[{{Fender}(2002)}]{fender:02a}
{Fender}, R. 2002, in LNP Vol. 589: Relativistic Flows in Astrophysics, 101

\bibitem[{{Gierli{\' n}ski} \& {Zdziarski}(2003)}]{gierlinski:03a}
{Gierli{\' n}ski}, M. \& {Zdziarski}, A.~A. 2003, \mnras, 343, L84

\bibitem[{{Gierlinski} {et~al.}(1998){Gierlinski}, {Zdziarski}, {Coppi},
    {Poutanen}, {Ebisawa}, \& {Johnson}}]{gierlinski:98a} {Gierlinski}, M.,
  {Zdziarski}, A.~A., {Coppi}, P.~S., {et~al.} 1998, in The Active X-ray
  Sky: Results from BeppoSAX and RXTE, ed. L.~Scarsi, H.~Bradt, P.~Giommi,
  \& F.~Fiore, Nuclear Physics B Proc.\ Supp. (Elsevier Science), 312

\bibitem[{{Gilfanov} {et~al.}(1999){Gilfanov}, {Churazov}, \&
  {Revnivtsev}}]{gilfanov:99a}
{Gilfanov}, M., {Churazov}, E., \& {Revnivtsev}, M. 1999, A\&A, 352, 182

\bibitem[{{Gleissner} {et~al.}(2003){Gleissner}, {Wilms}, {Pottschmidt},
  {Uttley}, {Nowak}, \& {Staubert}}]{gleissner:02a}
{Gleissner}, T., {Wilms}, J., {Pottschmidt}, K., {et~al.} 2003, in New Views on
  Microquasars, ed. P.~Durouchoux, Y.~Fuchs, \& J.~Rodriguez (Kolkata: Centre
  for Space Physics), 46

\bibitem[{Jahoda {et~al.}(1996)Jahoda, Swank, Giles, Stark, Strohmayer, Zhang,
  \& Morgan}]{jahoda:96}
Jahoda, K., Swank, J.~H., Giles, A.~B., {et~al.} 1996, in {EUV}, X-Ray, and
  Gamma-Ray Instrumentation for Astronomy {VII}, ed. O.~H. Siegmund, \& M.~A. Gummin, Proc.\
  SPIE Vol. 2808 (Bellingham, WA: SPIE), 59

\bibitem[{{Keeping}(1962)}]{keeping:62}
{Keeping}, E.~S. 1962, Introduction to statistical inference (Princeton: Van
  Nostrand)

\bibitem[{{Kotov} {et~al.}(2001){Kotov}, {Churazov}, \& {Gilfanov}}]{kotov:01a}
{Kotov}, O., {Churazov}, E., \& {Gilfanov}, M. 2001, MNRAS, 327, 799

\bibitem[{{Levine} {et~al.}(1996){Levine}, {Bradt}, {Cui}, {Jernigan},
  {Morgan}, {Remillard}, {Shirey}, \& {Smith}}]{levine:96}
{Levine}, A.~M., {Bradt}, H., {Cui}, W., {et~al.} 1996, \apjl, 469, L33

\bibitem[{Lochner {et~al.}(1991)Lochner, Swank, \& Szymkowiak}]{lochner:91}
Lochner, J.~C., Swank, J.~H., \& Szymkowiak, A.~E. 1991, ApJ, 376, 295

\bibitem[{{Lyubarskii}(1997)}]{lyubarskii:97a}
{Lyubarskii}, Y.~E. 1997, \mnras, 292, 679

\bibitem[{Maccarone \& Coppi(2002a)}]{maccarone:02a}
Maccarone, T.~J. \& Coppi, P.~S. 2002a, MNRAS, 335, 465

\bibitem[{{Maccarone} \& {Coppi}(2002b)}]{maccarone:02b}
{Maccarone}, T.~J. \& {Coppi}, P.~S. 2002b, \mnras, 336, 817

\bibitem[{Maccarone {et~al.}(2000)Maccarone, Coppi, \&
  Poutanen}]{maccarone:00a}
Maccarone, T.~J., Coppi, P.~S., \& Poutanen, J. 2000, ApJ, 537, L107

\bibitem[{{Mineshige} {et~al.}(1994){Mineshige}, {Takeuchi}, \&
  {Nishimori}}]{mineshige:94a}
{Mineshige}, S., {Takeuchi}, M., \& {Nishimori}, H. 1994, \apjl, 435, L125

\bibitem[{Mitsuda {et~al.}(1984)Mitsuda, Inoue, Koyama, Makishima, Matsuoka,
  Ogawara, Shibazaki, Suzuki, Yanaka, \& Hirano}]{mitsuda:84a}
Mitsuda, K., Inoue, H., Koyama, K., {et~al.} 1984, PASJ, 36, 741

\bibitem[{Miyamoto {et~al.}(1992)Miyamoto, Kitamoto, Iga, Negoro, \&
  Terada}]{miyamoto:92}
Miyamoto, S., Kitamoto, S., Iga, S., Negoro, H., \& Terada, K. 1992, ApJ, 391,
  L21

\bibitem[{Novikov \& Thorne(1973)}]{novikov:73a}
Novikov, I.~D. \& Thorne, K.~S. 1973, in Black Holes --- Les Astres Occlus, ed.
  C.~{DeWitt} \& B.~{DeWitt} (New York, London: Gordon and Breach), 345

\bibitem[{Nowak(2000)}]{nowak:00a}
Nowak, M.~A. 2000, MNRAS, 318, 361

\bibitem[{{Nowak}(2003)}]{nowak:02a}
{Nowak}, M.~A. 2003, in New Views on Microquasars, ed. P.~Durouchoux, Y.~Fuchs,
  \& J.~Rodriguez (Kolkata: Centre for Space Physics), 11

\bibitem[{{Nowak} {et~al.}(1998){Nowak}, {Dove}, {Vaughan}, {Wilms}, \&
  {Begelman}}]{nowak:98a}
{Nowak}, M.~A., {Dove}, J.~B., {Vaughan}, B.~A., {Wilms}, J., \& {Begelman},
  M.~C. 1998, in The Active X-ray
  Sky: Results from BeppoSAX and RXTE, ed. L.~Scarsi, H.~Bradt, P.~Giommi,
  \& F.~Fiore, Nuclear Physics B Proc.\ Supp. (Elsevier Science), 302

\bibitem[{Nowak {et~al.}(1999)Nowak, Vaughan, Wilms, Dove, \&
  Begelman}]{nowak:98b}
Nowak, M.~A., Vaughan, B.~A., Wilms, J., Dove, J.~B., \& Begelman, M.~C. 1999,
  ApJ, 510, 874

\bibitem[{{Nowak} {et~al.}(2002){Nowak}, {Wilms}, \& {Dove}}]{nowak:02b}
{Nowak}, M.~A., {Wilms}, J., \& {Dove}, J.~B. 2002, \mnras, 332, 856

\bibitem[{Pottschmidt {et~al.}(2000)Pottschmidt, Wilms, Nowak, Heindl, Smith,
  \& Staubert}]{pottschmidt:00a}
Pottschmidt, K., Wilms, J., Nowak, M.~A., {et~al.} 2000, A\&A, 357, L17

\bibitem[{Pottschmidt {et~al.}(2003)Pottschmidt, Wilms, Nowak, Pooley,
  Gleissner, Heindl, Smith, Remillard, \& Staubert}]{pottschmidt:03}
Pottschmidt, K., Wilms, J., Nowak, M.~A., {et~al.} 2003, A\&A, 407, 1039 (Paper I)

\bibitem[{Psaltis \& Norman(2001)}]{psaltis:00a}
Psaltis, D. \& Norman, C. 2001, ApJ, submitted (astro-ph/0001391)

\bibitem[{Scarsi {et~al.}(1998)Scarsi, Bradt, Giommi, \& Fiore}]{scarsi:98a}
Scarsi, L., Bradt, H., Giommi, P., \& Fiore, F., eds. 1998, The Active
{X-ray} Sky: Results from {BeppoSAX} and {RXTE}, Nuclear Physics B Proc.\ Supp.
  (Elsevier Science)

\bibitem[{Shakura \& Sunyaev(1973)}]{shakura:73}
Shakura, N.~I. \& Sunyaev, R.~A. 1973, A\&A, 24, 337

\bibitem[{Shapiro {et~al.}(1976)Shapiro, Lightman, \& Eardley}]{shapiro:76}
Shapiro, S.~L., Lightman, A.~P., \& Eardley, D.~M. 1976, ApJ, 204, 187

\bibitem[{{Smith} {et~al.}(2002){Smith}, {Heindl}, \& {Swank}}]{smith:01b}
{Smith}, D.~M., {Heindl}, W.~A., \& {Swank}, J.~H. 2002, \apj, 569, 362

\bibitem[{Sunyaev \& {Tr\"umper}(1979)}]{sunyaev:79a}
Sunyaev, R.~A. \& {Tr\"umper}, J. 1979, Nature, 279, 506

\bibitem[{{Thorne} \& {Price}(1975)}]{thorne:75}
{Thorne}, K.~S. \& {Price}, R.~H. 1975, \apjl, 195, L101

\bibitem[{{Uttley} \& {McHardy}(2001)}]{uttley:01}
{Uttley}, P. \& {McHardy}, I.~M. 2001, \mnras, 323, L26

\bibitem[{{van der Klis}(1989)}]{klis:89a}
{van der Klis}, M. 1989, in Timing Neutron Stars, ed. H.~\"Ogelman \& E.~P.~J.
  {van den Heuvel}, NATO ASI No. C262 (Dordrecht: Kluwer Academic Publishers),
  27

\bibitem[{{van der Klis}(1995)}]{klis:95}
{van der Klis}, M. 1995, in {X-ray} Binaries, ed. W.~H.~G. Lewin, J.~{van
  Paradijs}, \& E.~P.~J. {van den Heuvel} (Cambridge: Cambridge University Press),
  252

\bibitem[{{van der Klis}(2001)}]{klis:01a}
{van der Klis}, M. 2001, \apj, 561, 943

\bibitem[{{van Straaten} {et~al.}(2003){van Straaten}, {van der Klis}, \&
  {M\'endez}}]{vanstraaten:03a}
{van Straaten}, S., {van der Klis}, M., \& {M\'endez}, M. 2003, \apj, 596, 1155

\bibitem[{{Witt} {et~al.}(1997){Witt}, {Czerny}, \& {\.Zycki}}]{witt:97}
{Witt}, H.~J., {Czerny}, B., \& {\.Zycki}, P.~T. 1997, \mnras, 286, 848

\bibitem[{{Zdziarski} {et~al.}(2002){Zdziarski}, {Poutanen}, {Paciesas}, \&
  {Wen}}]{zdziarski:02}
{Zdziarski}, A.~A., {Poutanen}, J., {Paciesas}, W.~S., \& {Wen}, L. 2002, \apj,
  578, 357

\bibitem[{\.Zycki(2003)}]{zycki:03a}
\.Zycki, P.~T. 2003, MNRAS, 340, 639

\end{thebibliography}
\end{document}